# FROM UNBALANCED INITIAL OCCUPANT DISTRIBUTION TO BALANCED EXIT USAGE IN A SIMULATION MODEL OF PEDESTRIAN DYNAMICS

Tobias Kretz and Andree Große
PTV Group, Haid-und-Neu-Str. 15, D-76131 Karlsruhe, Germany
`Tobias.Kretz@ptvgroup.com`

**ABSTRACT**

It is tested in this contribution if and to which extend a method of a pedestrian simulation tool that attempts to make pedestrians walk into the direction of estimated earliest arrival can help to automatically distribute pedestrians – who are initially distributed arbitrarily in the scenario – equally on the various exits of the scenario.

## INTRODUCTION

"Everyone walking to the nearest exit" is one of the exit selection paradigms which are imagined to be applied by occupants a building in an emergency situation (another one which is often mentioned is "everyone through the exit which he/she used as entrance). This is odd in a way as there is no direct gain in minimizing walking distance in case of emergency. The gain only exists with the implicit assumption that shortest travel distance implies smallest travel time which reduces exposure time to smoke or time spent in a risky environment (e.g. a building destabilized by an earth quake). This, however, is not necessarily true if occupants get stuck in jams where for them time passes by while they make only small (spatial) progress. Here there is another implicit equation which is true more often: higher densities along the path imply higher travel times. As the actual intention of evacuation procedures is to reduce travel times, occupants ought to assess densities along the paths to various exits together with the distances to these exits and then decide for an exit.

The reason why a planner would want to simulate the system at user equilibrium is not necessarily that he assumes that the occupants would actually themselves distribute like this (although the assumption can be justified), but it may be just to learn about the minimum evacuation time which could be achieved to have a fix point against which actual planning can be assessed or to learn from which regions occupants need to be sent to which exit to realize user equilibrium by the external organizing hand of the planner.

## ESTIMATING REMAINING TRAVEL TIME

Traditionally models of pedestrian dynamics explicitly or implicitly shared the assumption that pedestrians strictly prefer to walk on the shortest path (Schadschneider, et al., 2009). Recently a method has been introduced in which the spatial distribution of pedestrians and their walking state is used as input to a numerical Eikonal equation solver to compute a map – in the form of a regular grid with grid point spacing below body radius – of estimated remaining travel time for a particular destination. Except for the role which the two investigated parameters have (see below), the details of the method are not of much relevance here and can be found elsewhere (Kretz, et al., 2011). Important here is that the map covers all of the space which is accessible by pedestrians and that implicitly with the (negative) gradients implicitly the continuously updated map gives an (estimated) quickest path to the destination with the restriction that simulated just as real pedestrians normally are far from having all information available which would be necessary to calculate the actual quickest path under given circumstances. Considering estimated time delays for the choice of the preferred walking direction in a simulation model of pedestrian dynamics has major impacts in many situations (Kretz, 2009) (Kretz, 2009b) (Kretz, 2011) (Kretz, 2012)

The method is mainly controlled by two parameters called $g$ and $h$. The value of parameter $g$ sets the basic amount of estimated delay which is introduced by the presence of a pedestrian for pedestrians being upstream. The word "upstream" is not strictly defined in the two-dimensional movement environment of pedestrians, still it describes the relation best as for most of the area someone can intuitively say if it up- or downstream of a position with regard to a certain destination. If parameter $g=0$, then the map of estimated remaining travel times is proportional to the map of distances by a global factor. Depending on the grid point spacing of the map it can be shown that at values between $g=1.5$ and $g=4.0$ that impact on the map which comes from a specific pedestrian X will start to have an increasingly large impact on that same pedestrian X in the next simulation time step. This is highly undesired as pedestrians then will start to move "nervously". The range of useful values for parameter $g$ therefore is limited to values of about $g=5.0$. Parameter $h$ sets the relevance of the movement direction of a pedestrian for its impact on the map of estimated remaining travel times. If $h=0$ then it is irrelevant, where a pedestrian is walking to. For larger values the impact on estimated travel times is larger, when a pedestrian comes across, i.e. walks in upstream direction with regard of the destination for which a map is calculated. With $h=1.0$ the impact vanishes approximately, when a pedestrian walks with free speed towards the destination. Normally $h=1.0$ will therefore be the upper bound of values for parameter $h$.

## TEST SCENARIO AND SIMULATION SETUP

In a highly symmetric scenario with symmetric geometry, symmetric exit positions, homogeneous walking speeds, and symmetric initial occupant distribution a balanced distribution of occupants on exits can result automatically (if the occupants are aware of all exits which is assumed in this contribution) with a nearest exit approach. However, if one or more of these conditions are not met a nearest exit approach can quickly yield exit usage far off user equilibrium. In this contribution the case of unbalanced initial occupant distribution within a symmetric geometry is investigated. For this the geometry of test case 9 of IMO's MSC/Circ. 1238 (International Maritime Organization, 2007) (which is identical to the test with the same number from RiMEA (Brunner, et al., 2009)) is chosen, while the initial distribution of pedestrians is altered compared to the standard test case, see Figure 1[1].

---

[1] Note that RiMEA's test case11 is an example where there are two exits at different distances and that in (Kretz, 2010) an example is investigated where there are two exits at equal distance but with different capacity/width.

The scenario was simulated with Viswalk (PTV Group, 2011). Viswalk utilizes the Social Force Model (Johansson, et al., 2007) and includes aforementioned method for remaining travel time estimation to make pedestrians set the direction of estimated least remaining travel time as preferred walking direction.

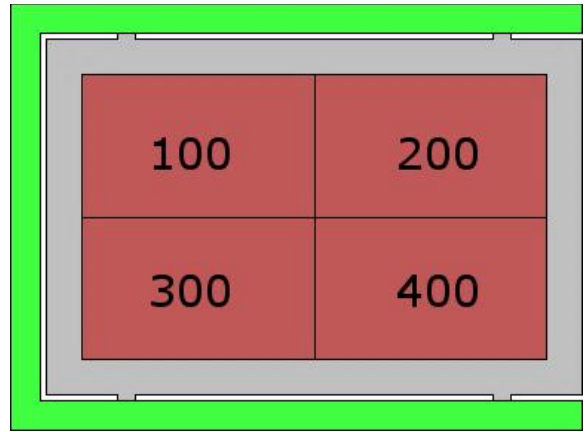

Figure 1: *Modified initial spatial pedestrian distribution of IMO's test case 9. Instead of placing initially 250 occupants in each of the four quadrants (the red / dark grey areas in the middle), the initial distribution is chosen to be as shown by the numbers. As the quadrants are areas which by distance can be assigned to one of the four exits, the distribution that results from "everyone to the nearest exit" would be unbalanced and evacuation time would be considerably higher than in the case of balanced initial density. With the surrounding green / dark grey area all four exits are combined to be one, such that there is just one single travel time assessment and the simulated pedestrians implicitly and automatically turn to the right exit (the one which promises smallest estimated individual travel time)*

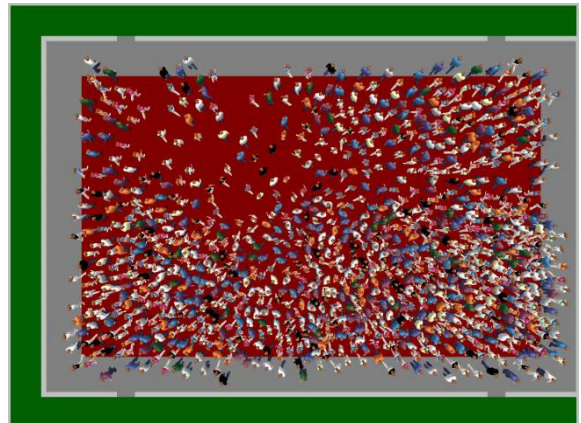

Figure 2: *Situation at beginning (one second after simulation start).*

## RESULTS

Table 1 shows how many pedestrians – on an average of 100 simulations – walked through each of the four exits. It can be seen that even setting the direction of the shortest path as preferred walking direction ($g=0.0$) does not make every pedestrian walk through the nearest exit (the exit usage is different from the initial distribution). The reason for this is that the repelling forces between pedestrians because of the high density on the lower right area overrule the preferred walking direction and lead to a somewhat more balanced distribution to the exits than is provided by the initial distribution.

With increasing value of parameter *g* the distribution on the exits approaches more and more an equal distribution. As Table 2 shows this has the effect that the total evacuation time is reduced by a factor close to the reduction of the number of pedestrians using the most frequented exit at the lower right. The average of individual egress times is reduced as well, although – unsurprisingly – to a lesser degree.

As in this scenario soon after the simulation start all pedestrians stand about still in jams in front of doors, the value of parameter *h* only has a minor influence on the results in this case. This was verified by simulating with values *h=0.0* and *h=1.0* and parameter g fixed to *g=2.0*.

Table 1: *Average (of 100 simulation runs) number of pedestrians using an exit.*

|  | Upper left | Upper right | Lower left | Lower right |
|---|---|---|---|---|
| *Initial (for comparison)* | *100* | *200* | *300* | *400* |
| **Shortest Path (*g=0.0*)** | 156.3 | 253.4 | 269.5 | 320.8 |
| **Quickest Path (*g=1.0, h=0.0*)** | 206.6 | 252.7 | 249.7 | 291.1 |
| **Quickest Path (*g=2.0, h=0.0*)** | 219.0 | 251.8 | 247.9 | 281.3 |
| **Quickest Path (*g=2.0, h=1.0*)** | 216.7 | 250.4 | 252.2 | 280.7 |
| **Quickest Path (*g=3.0, h=0.0*)** | 224.5 | 252.0 | 248.4 | 275.1 |
| **Quickest Path (*g=5.0, h=0.0*)** | 230.8 | 250.7 | 249.4 | 269.2 |

Table 2: *Evacuation times (average of individual egress times and total evacuation time).*

|  | Individual Av. [s] | Total (Last) [s] |
|---|---|---|
| **Equal (Balanced) Initial Distribution (4 x 250)** |  |  |
| **Shortest Path (*g=0.0*)** | 81.6 | 202.1 |
| **Quickest Path (*g=1.0, h=0.0*)** | 76.3 | 178.7 |
| **Quickest Path (*g=2.0, h=0.0*)** | 75.3 | 171.7 |
| **Quickest Path (*g=2.0, h=1.0*)** | 75.7 | 173.1 |
| **Quickest Path (*g=3.0, h=0.0*)** | 75.0 | 167.6 |
| **Quickest Path (*g=5.0, h=0.0*)** | 74.9 | 164.0 |

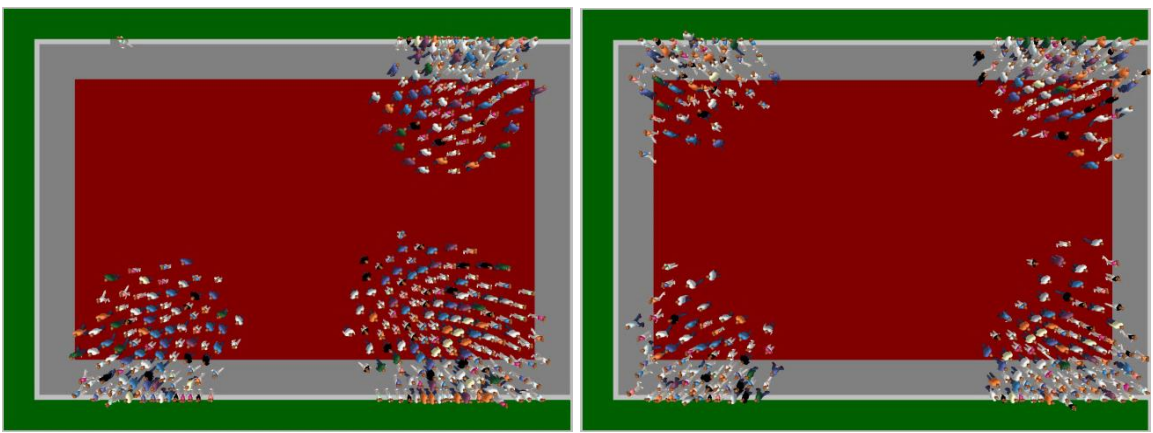

Figure 3: *With g=0.0 (left) and g=5.0 (right) after 100 seconds.*

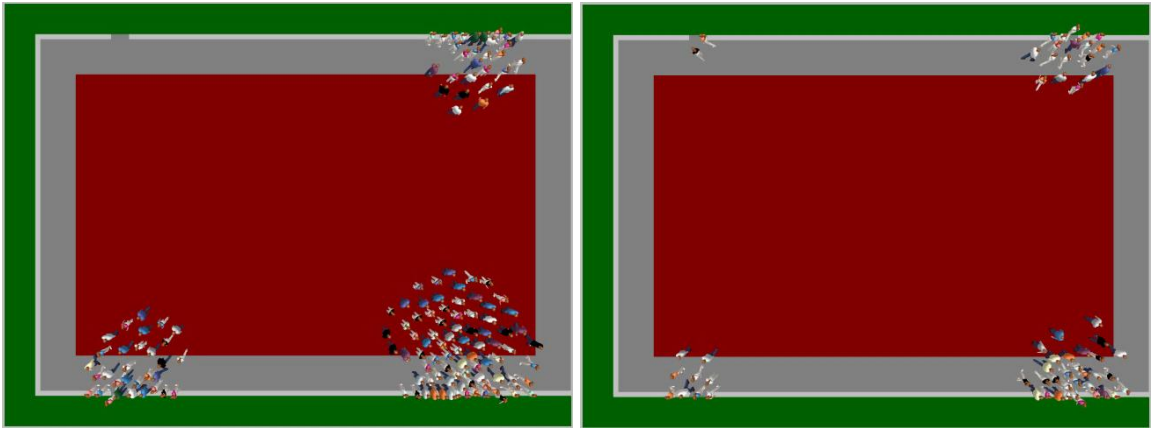

Figure 4: *With g=0.0 (left) and g=5.0 (right) after 140 seconds.*

## CONCLUSIONS

It was shown for a simple walking geometry for which the optimal solution of distribution of occupants on the four available exits is trivially known that a method that assigns the direction of estimated least remaining travel time (estimated quickest arrival) as preferred walking direction is able to largely compensate an inhomogeneous spatial distribution of pedestrians and come up with a balanced distribution of pedestrians on the exits. This has a positive effect on the total evacuation time as well as the average individual egress time.

How likely it is that real people as occupants in a case of emergency are able to automatically distribute so equally (and therefore optimally) on various exits surely depends on the specific case. A method as the one probed in this contribution therefore does not necessarily predict how the procedure will evolve by itself, but it indicates which measures need to be implemented to effectuate this behavior.